\documentclass[conference]{IEEEtran}
\IEEEoverridecommandlockouts
\usepackage{cite}
\usepackage[tbtags]{amsmath}
\usepackage{amssymb,amsfonts}
\usepackage{algorithm}
\usepackage{algorithmic}
\usepackage{graphicx}
\usepackage{textcomp}
\usepackage{xcolor}
\usepackage{caption}
\usepackage{commath}
\usepackage{url}

\topmargin \dimexpr 1in -1.721in

\def\BibTeX{{\rm B\kern-.05em{\sc i\kern-.025em b}\kern-.08em
    T\kern-.1667em\lower.7ex\hbox{E}\kern-.125emX}}

\begin{document}
\title{Low-Complexity SDP-ADMM for Physical-Layer Multicasting in Massive MIMO Systems}
\author{\thanks{This work was supported by the FFL18-0277 grant from the Swedish Foundation for Strategic Research.}
\IEEEauthorblockA{Mahmoud Zaher, and Emil Björnson\\
 \{mahmoudz, emilbjo\}@kth.se\\}

\IEEEauthorblockA{
Department of Computer Science, KTH Royal Institute of Technology, Sweden
}
}
\maketitle

\begin{abstract}

There is a demand for the same data content from several user equipments (UEs) in many wireless communication applications. Physical-layer multicasting combines the beamforming capability of massive MIMO (multiple-input multiple-output) and the broadcast nature of the wireless channel to efficiently deliver the same data to a group of UEs using a single transmission. This paper tackles the max-min fair (MMF) multicast beamforming optimization, which is an NP-hard problem. We develop an efficient semidefinite program-alternating direction method of multipliers (SDP-ADMM) algorithm to find the near-global optimal rank-1 solution to the MMF multicast problem in a massive MIMO system. Numerical results show that the proposed SDP-ADMM algorithm exhibits similar spectral efficiency performance to state-of-the-art algorithms running on standard SDP solvers at a vastly reduced computational complexity. We highlight that the proposed ADMM elimination procedure can be employed as an effective low-complexity rank reduction method for other problems utilizing semidefinite relaxation.

\end{abstract}


\section{Introduction}

Massive MIMO (multiple-input multiple-output) technology has emerged as a key enabler for wireless communication systems, offering significant gains in spectral efficiency (SE) \cite{bjornson2017book,zaher2023learning}. By deploying many antennas at the base station (BS), massive MIMO facilitates effective beamforming and spatial multiplexing, which can greatly boost the desired signal quality of multiple user equipments (UEs). Through beamforming, physical-layer multicasting can simultaneously deliver the same information to multiple UEs within a coverage area using a single transmission. This approach mitigates the problem of co-channel interference and is particularly beneficial to efficiently utilize network resources \cite{hsu2016joint}.

Physical-layer multicasting can support a wide range of applications in modern wireless communication networks. For instance, mobile network operators can leverage it for video streaming of live events to multiple subscribers, providing a cost-effective alternative to traditional unicast-based content delivery. Other applications of multicasting include videoconferencing as well as broadcasting of machine learning models in federated learning applications, where multiple UEs need to receive the same data stream. Another use case is emergency broadcast systems where critical information needs to be disseminated rapidly to a large audience, ensuring efficient and timely communication.

A foundational framework for physical-layer multicasting is presented in 
\cite{sidiropoulos2006transmit}, introducing quality of service (QoS) and max-min fair (MMF) designs for a single-cell single-group multicast scenario, where the QoS level ensures a minimum received SNR for all UEs, and MMF aims to maximize the lowest received SNR. This concept was later expanded in \cite{karipidis2008quality} to include multiple co-channel multicast groups. The problem was proven to be NP-hard, and approximate solutions were derived using semidefinite relaxation (SDR). In \cite{karipidis2008quality,sidiropoulos2006transmit}, different randomization strategies were proposed to extract near-optimal rank-1 beamforming solutions from higher-rank semidefinite program (SDP) output. However, as the number of transmit antennas and UEs grow, the approximation accuracy deteriorates significantly \cite{sidiropoulos2006transmit,karipidis2008quality,zhou2017coordinated}. Note that a rank-1 beamforming solution, where the BS transmits a single multicast signal using spatial multiplexing, represents a practical choice due to its ease of implementation compared to a higher-rank transmission.

Recent studies have introduced different iterative optimization techniques to address the QoS and MMF multicast problems. In \cite{zhou2017coordinated}, the authors formulate a coordinated MMF multicast problem in a multicell network, where each BS serves a single UE group, and solve it using parametric manifold optimization. The study in \cite{hsu2016joint} proposes three different algorithms for the MMF problem under per-cell power constraints with joint beamforming across cooperating BSs. Among these, the difference-of-convex approximation (DCA) algorithm demonstrates superior SE while achieving lower computational complexity compared to traditional SDR methods with randomization. The DCA algorithm is built upon the successive convex approximation (SCA) framework, which represents the state-of-the-art in multicast beamforming optimization. A two-layer beamforming scheme is introduced in \cite{sadeghi2017reducing}, where the first layer suppresses multicast inter-group interference, and the second layer utilizes SCA to maximize the SNR of the decoupled multicast groups. In \cite{your_globecom_paper}, we propose a successive elimination algorithm (SEA) that relies on SDR followed by iterative elimination of higher-rank solutions to extract a near-optimal rank-1 solution to the MMF multicast problem.
Additionally, the authors in \cite{dong2020multi} present alternative formulations for the QoS and MMF problems, leveraging the optimal multicast beamforming structure, which provides a reduction in computational complexity when the number of BS antennas is much greater than the number of UEs.

A common approach in the prior literature \cite{chen2017admm,konar2017fast,zhang2023ultra} is to utilize the SCA technique together with the alternating direction method of multipliers (ADMM) to achieve a low-complexity local optimum solution to the QoS and MMF multicast problems. It is worth noting that, when general-purpose solvers were used to solve the optimization problems, our previously proposed SEA has shown superior performance in terms of SE and computational complexity compared to state-of-the-art algorithms that rely on SCA \cite{your_globecom_paper}. A detailed performance evaluation and comparison of the proposed algorithm to the SCA-ADMM based algorithms will be provided in the extended journal article of this paper.

The advancements in multicast beamforming research highlight the increasing importance of efficient transmission techniques for emerging wireless applications. In this paper, we propose a novel SDP-ADMM algorithm to solve the MMF multicast problem in massive MIMO systems. Notably, this represents the first application of SDP-ADMM in multicast beamforming optimization, demonstrating its strong potential for improving efficiency in this field. We develop a new problem reformulation and ADMM-based implementation, extending our previously proposed SEA framework in \cite{your_globecom_paper}, to obtain a near-global optimal rank-1 beamforming solution. A key advantage of our approach is its adaptability to various multicast optimization objectives and network configurations without requiring ADMM parameter tuning. Throughout the paper, we focus on multicast beamforming design, assuming perfect CSI at the transmitting BS. Our results confirm that the proposed SDP-ADMM algorithm achieves comparable SE performance to the SEA that relies on conventional SDP solvers, while significantly reducing computational complexity.

\section{System Model} \label{sys_mod}

\begin{figure}
\centering
\setlength{\abovecaptionskip}{0.33cm plus 0pt minus 0pt}
\includegraphics[scale=0.65]{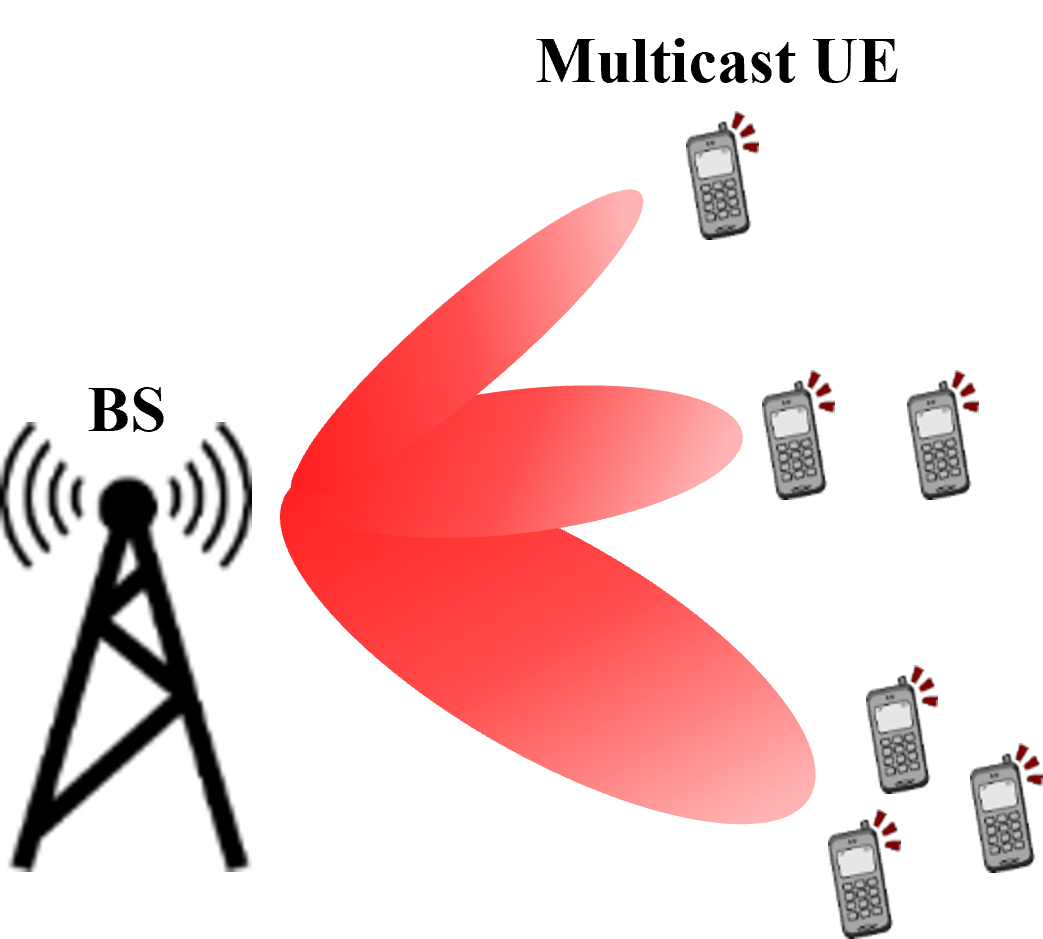}
\caption{Massive MIMO multicasting with irregular beamforming to a UE group.}
\label{system_model_fig}
\vspace{-0.65em}
\end{figure}

This paper considers a massive MIMO BS equipped with $N$ antennas. The BS serves $K$ single-antenna UEs, that are arbitrarily distributed in a large service area, with a single multicast transmission. We consider a narrow-band channel, such that each channel realization is frequency-flat and quasi-static in time. The channel realizations are assumed to be available at the BS.
The channel between UE $k$ and the BS is denoted as $\mathbf{h}_{k} \in \mathbb{C}^N$. The system model is depicted in Fig. \ref{system_model_fig}.

The received signal at UE $k$ is computed as
\begin{equation}
    y_{k}^\mathrm{dl} = \mathbf{h}_{k}^H\mathbf{w}s + n_{k},
\end{equation}
where $s$ denotes the zero-mean unit-variance multicast signal intended for all UEs, $\mathbf{w}$ represents the common multicast precoding vector, and $n_{k} \sim \mathcal{N}_{\mathbb{C}}(0, \sigma_{k}^2)$ is the noise at UE $k$. The multicast precoding vector satisfies a short-term power constraint, which means that the power constraint must be satisfied for each channel realization. Accordingly, $\norm{\mathbf{w}}^2 \leq P_{\mathrm{max}}$, where $P_{\mathrm{max}}$ represents the maximum transmit power of the BS. As a result, an achievable SE of the $k^{\textrm{th}}$ UE under the perfect CSI assumption can be evaluated as
\begin{equation}
    \textrm{SE}_{k}^{\mathrm{dl}} = \textrm{log}_2\left(1 + \textrm{SNR}_k^{\mathrm{dl}}\right).
    \label{SE}
\end{equation}

We define the notation $\mathbf{H}_k = \mathbf{h}_k\mathbf{h}_k^H/\sigma_k^2$ and $\mathbf{W} = \mathbf{w}\mathbf{w}^H$. Utilizing the fact that $\left|\mathbf{h}_{k}^H\mathbf{w}\right|^2 = \textrm{tr}\left(\mathbf{h}_{k}\mathbf{h}_{k}^H\mathbf{w}\mathbf{w}^H\right)$, the SNR of UE $k$ in \eqref{SE} can be written as
\begin{equation}
    \textrm{SNR}_k^{\mathrm{dl}} = \frac{\left|\mathbf{h}_{k}^H\mathbf{w}\right|^2}{\sigma_{k}^2} = \textrm{tr}\left(\mathbf{H}_k\mathbf{W}\right).
    \label{SNR}
\end{equation}

This SNR formulation is suitable to construct the SDP of the MMF multicast problem.

\subsection{Max-Min Fair Multicast Problem} \label{max_min_sec}

In this paper, the primary focus is on the MMF multicast problem. The goal is to find the multicast transmit precoding vector $\mathbf{w}$ that maximizes the minimum achievable SNR in \eqref{SNR} among all UEs in the system under the power constraint at the BS. Since all UEs in a multicast transmission must be able to decode the same signal, it is important to note that the minimum SNR among the UEs determines what data rate can be utilized. Hence, an appropriate design criterion for the multicast problem is to maximize the minimum achievable SNR.

We utilize the relation between the QoS and MMF multicast problems to solve the MMF problem. The logic is that the target SNR of the QoS problem represents the MMF SNR that is achievable given a power budget at the BS equal to the optimum objective value achieved for the QoS problem. Since the optimum objective values of the MMF and QoS problems are monotonically non-decreasing in the BS power budget and the SNR target, respectively, a bisection search can be performed over the SNR target to identify the max-min SNR that results in an optimum objective value of the QoS problem equal to the power budget available at the BS.
This relation is well-established in the previous literature, e.g., \cite{sidiropoulos2006transmit,chen2017admm,your_globecom_paper}.

Let $\mathbb{H}^N$ denote the set of $N \times N$ Hermitian matrices. The linear map $\mathcal{H}\left(\cdot\right): \mathbb{H}^N \rightarrow \mathbb{R}^K$ is defined as
\begin{equation}
    \mathcal{H}\left(\mathbf{W}\right) = \left(\langle\mathbf{H}_1^H, \mathbf{W}\rangle, \hdots, \langle\mathbf{H}_K^H, \mathbf{W}\rangle\right),
\end{equation}
where the inner product between two matrices is given by $\langle\mathbf{A}, \mathbf{B}\rangle = \textrm{tr}\left(\mathbf{A}^H\mathbf{B}\right)$. The QoS problem can then be written compactly as
\begin{subequations}
\begin{align}
    \min_{\mathbf{W}} \quad &\textrm{tr}\left(\mathbf{W}\right) \\
    \textrm{s.t.}\quad&\mathcal{H}\left(\mathbf{W}\right) \geq \boldsymbol{\gamma},\\
    &\mathbf{W} \succeq \mathbf{0}, \quad \textrm{rank}\left(\mathbf{W}\right) = 1,\label{rank1}
\end{align}
\label{qos_problem}%
\end{subequations}
where $\boldsymbol{\gamma} \in \mathbb{R}^{K}$ represents the SNR targets of the UEs. The above problem is non-convex due to the non-convex rank-1 constraint in \eqref{rank1}. A relaxation of the problem is devised by dropping the rank-1 constraint to obtain the relaxed convex QoS problem. In general, such relaxation results in the solution matrix $\mathbf{W}$ having a high rank that does not satisfy the rank constraint of the original problem, requiring post-processing of the output matrix to extract a feasible rank-1 beamforming solution. Finding the optimal post-processing is challenging.

\subsection{Successive Elimination Algorithm}

Low-rank matrix solutions to SDP problems belong to the boundary of the feasible set and not the interior. This means that the solution to the relaxed SDP problem will most likely have a high rank, even if a rank-1 solution exists. In fact, standard SDP solvers will always favor a higher-rank solution over the lower-rank counterpart, even if both can achieve the same objective value \cite{luo2010semidefinite}. Previous works relying on SDR have generally used a randomization procedure \cite{karipidis2008quality} to extract a rank-1 beamforming solution. However, this method does not scale well and a large number of candidate sets must be tested when increasing the number of BS antennas and UEs for satisfactory SE performance. For that purpose, we utilize the state-of-the-art SEA proposed in our primal work \cite{your_globecom_paper}. The idea is to iteratively eliminate the output higher-rank solutions to the SDP that result in an optimum objective value greater than or equal to that achieved with the optimum rank-1 solution, until a near-optimal rank-1 solution is achieved. The elimination is done through penalizing the directions of the eigenvectors corresponding to the second-largest eigenvalues of the higher-rank solutions of the previous iterations. The penalty is in the form of a quadratic matrix product $\zeta\sum_{r}\mathbf{u}_r^H\mathbf{W}\mathbf{u}_r$, where $\zeta \geq 1$ represents the penalty factor and the $\mathbf{u}_r$, $\forall r$, represent the eigenvectors corresponding to the second-largest eigenvalues of the previous higher-rank solutions.

In this paper, we propose a new formulation that is suitable for developing a computationally fast and effective ADMM algorithm to solve the QoS problem. The penalized relaxed QoS problem is formulated as
\begin{subequations}
\begin{align}
    \min_{\mathbf{W}} \quad &\langle\boldsymbol{\Lambda}, \mathbf{W}\rangle \\
    \textrm{s.t.} \quad &\mathcal{H}\left(\mathbf{W}\right) \geq \boldsymbol{\gamma},\label{SNR_constraint}\\
    &\mathbf{W} \succeq \mathbf{0},\label{psd_constraint}
\end{align}
\label{penalized_qos_problem}%
\end{subequations}
where $\boldsymbol{\Lambda} \in \mathbb{C}^{N \times N}$ is a design parameter that is responsible for the elimination of the higher-rank solutions and improves the convergence speed of the proposed algorithm, and will be detailed later.

\section{SDP-ADMM: Efficient Multicast Beamforming}

In this section, we develop an efficient SDP-ADMM algorithm to solve the MMF multicast beamforming problem. The proposed algorithm relies on ADMM and utilizes the relation between the MMF and QoS problems as well as the SEA to find the near-optimal rank-1 MMF beamforming vectors.

\subsection{A Fast SDP-ADMM Algorithm For The QoS Problem}\label{SDP_ADMM}

Since the penalized relaxed QoS problem in \eqref{penalized_qos_problem} is convex and satisfies Slater's condition, strong duality holds and it is easier to solve the dual SDP problem. We begin by constructing the Lagrangian of problem \eqref{penalized_qos_problem} as
\begin{equation}
    L_q\left(\mathbf{W}, \mathbf{y}, \mathbf{S}\right) = \langle\boldsymbol{\Lambda}, \mathbf{W}\rangle - \langle \mathbf{y}, \mathcal{H}\left(\mathbf{W}\right) - \boldsymbol{\gamma}\rangle - \langle\mathbf{S}, \mathbf{W}\rangle,
    \label{lagrangian}%
\end{equation}
where $\mathbf{y} \in \mathbb{R}^K_+$ and $\mathbf{S} \in \mathbb{H}^N_+$ represent the nonnegative and positive semidefinite Lagrange multipliers of the constraints in \eqref{SNR_constraint} and \eqref{psd_constraint}, respectively. The adjoint operator of $\mathcal{H}$ is $\mathcal{H}^H\left(\cdot\right): \mathbb{R}^K \rightarrow \mathbb{H}^N$, and is defined as $\mathcal{H}^H\left(\mathbf{y}\right) = \sum_{k = 1}^Ky_k\mathbf{H}_k$. Taking the derivative of the Lagrangian with respect to $\mathbf{W}$ and equating it to $\mathbf{0}$, we get

\begin{equation}
    \mathcal{H}^H\left(\mathbf{y}\right) + \mathbf{S} = \boldsymbol{\Lambda}.
    \label{condition}
\end{equation}

Plugging this condition back in \eqref{lagrangian}, the dual function becomes $g\left(\mathbf{W}, \mathbf{y}, \mathbf{S}\right) = \mathbf{y}^T\boldsymbol{\gamma}$ when the condition in \eqref{condition} is satisfied. The dual SDP problem can then be formulated as
\begin{subequations}
\begin{align}
    \min_{\mathbf{y}, \mathbf{S}} \quad &\hspace{-3pt}-\mathbf{y}^T\boldsymbol{\gamma}\\
    \textrm{s.t.} \quad &\mathcal{H}^H\left(\mathbf{y}\right) + \mathbf{S} = \boldsymbol{\Lambda},\\
    &\mathbf{S} \succeq \mathbf{0},\quad \mathbf{y}\geq \mathbf{0}.
\end{align}
\label{dual_problem}%
\end{subequations}

We define the indicator function $\mathbb{I}_{\mathbf{y} \geq \mathbf{0}}\left(\mathbf{y}\right)$ as
\begin{equation}
    \mathbb{I}_{\mathbf{y} \geq \mathbf{0}}\left(\mathbf{y}\right) =
    \begin{cases}
        0 &\textrm{for } \mathbf{y} \geq \mathbf{0}, \\
        +\infty &\textrm{otherwise.}
    \end{cases}
\end{equation}
Then, an equivalent reformulation of the dual SDP problem is
\begin{subequations}
\begin{align}
    \min_{\mathbf{y}, \mathbf{S}} \quad &\hspace{-3pt}-\mathbf{y}^T\boldsymbol{\gamma} + \mathbb{I}_{\mathbf{y} \geq \mathbf{0}}\left(\mathbf{y}\right)\\
    \textrm{s.t.} \quad &\mathcal{H}^H\left(\mathbf{y}\right) + \mathbf{S} = \boldsymbol{\Lambda},\\
    &\mathbf{S} \succeq \mathbf{0}.
\end{align}
\label{dual_problem2}%
\end{subequations}
The augmented Lagrangian, in scaled form, for the dual SDP corresponding to the linear constraints can thus be written as
\begin{equation}
\begin{split}
    L_{\rho}&\left(\mathbf{W},\mathbf{y},\mathbf{S}\right) = -\mathbf{y}^T\boldsymbol{\gamma} + \mathbb{I}_{\mathbf{y} \geq \mathbf{0}}\left(\mathbf{y}\right) \\
    &+ \frac{\rho}{2}\norm{\mathcal{H}^H\left(\mathbf{y}\right) + \mathbf{S} - \boldsymbol{\Lambda} + \mathbf{W}}_F^2.
\end{split}
\end{equation}

Utilizing the ADMM algorithm \cite{boyd2011distributed}, the minimization is done with respect to each block of variables separately while the other variables are kept fixed. At each iteration, the following updates are computed sequentially:
\begin{align}
    &\mathbf{y}^{i+1} = \mathrm{arg\,}\mathop{\mathrm{min}}\limits_{\mathbf{y}}\,L_{\rho}\left(\mathbf{W}^i,\mathbf{y},\mathbf{S}^i\right),\label{y_update}\\
    &\mathbf{S}^{i+1} = \mathrm{arg\,}\mathop{\mathrm{min}}\limits_{\mathbf{S}}\,L_{\rho}\left(\mathbf{W}^i,\mathbf{y}^{i+1},\mathbf{S}\right), \textrm{ s.t. } \mathbf{S} \succeq \mathbf{0}, \label{S_update}\\
    &\mathbf{W}^{i+1} = \mathbf{W}^i + \mathcal{H}^H\left(\mathbf{y}^{i+1}\right) + \mathbf{S}^{i+1} - \boldsymbol{\Lambda}.
\end{align}

\subsubsection{$\mathbf{y}$-Update}

Introducing an auxiliary variable $\mathbf{z}$ and the indicator function $\mathbb{I}_{\mathbf{z} \geq \mathbf{0}}\left(\mathbf{z}\right)$, the update of $\mathbf{y}$ in problem \eqref{y_update} is equivalent to solving the following problem:
\begin{subequations}
\begin{align}
    \min_{\mathbf{y},\mathbf{z}} \quad &\hspace{-3pt}-\mathbf{y}^T\boldsymbol{\gamma} + \frac{\rho}{2}\norm{\mathcal{H}^H\left(\mathbf{y}\right) + \mathbf{S}^i - \boldsymbol{\Lambda} + \mathbf{W}^i}_F^2 + \mathbb{I}_{\mathbf{z} \geq \mathbf{0}}\left(\mathbf{z}\right)\\
    \textrm{s.t.} \quad & \mathbf{y} = \mathbf{z}.
\end{align}
\label{y_update_problem}
\end{subequations}

This problem can be efficiently solved using an (inner) ADMM algorithm to find the optimal $\mathbf{y}$-update. The augmented Lagrangian of problem \eqref{y_update_problem} can be formulated with the scaled dual variable $\mathbf{g}$ as
\begin{equation}
\begin{split}
    L_{\rho,\mu}\hspace{-2pt}\left(\mathbf{W}^i,\mathbf{y},\mathbf{S}^i,\mathbf{z}, \mathbf{g}\right) &= -\mathbf{y}^T\boldsymbol{\gamma} + \frac{\mu}{2}\norm{\mathbf{y} - \mathbf{z} + \mathbf{g}}_2^2 + \mathbb{I}_{\mathbf{z} \geq \mathbf{0}}\left(\mathbf{z}\right) \\
    &+ \frac{\rho}{2}\norm{\mathcal{H}^H\left(\mathbf{y}\right) + \mathbf{S}^i - \boldsymbol{\Lambda} + \mathbf{W}^i}_F^2.
\end{split}
\label{y_update_lagrange}
\end{equation}

The update of the variable blocks is done successively using the ADMM algorithm, while other variables are kept fixed. Accordingly, the inner ADMM iterations require computing the following updates:
\begin{align}
    &\mathbf{y}^{j+1} = \mathrm{arg\,}\mathop{\mathrm{min}}\limits_{\mathbf{y}}\,L_{\rho, \mu}\left(\mathbf{W}^i,\mathbf{y},\mathbf{S}^i, \mathbf{z}^j, \mathbf{g}^j\right),\\
    &\mathbf{z}^{j+1} = \mathrm{arg\,}\mathop{\mathrm{min}}\limits_{\mathbf{z}}\,L_{\rho, \mu}\left(\mathbf{W}^i,\mathbf{y}^{j+1},\mathbf{S}^i, \mathbf{z}, \mathbf{g}^j\right),\label{z_update}\\
    &\mathbf{g}^{j+1} = \mathbf{g}^j + \mathbf{y}^{j+1} - \mathbf{z}^{j+1}. \label{g_update}
\end{align}

Note that the iteration index for the outer ADMM algorithm is omitted for the variable $\mathbf{y}$ to simplify the notation. For the inner $\mathbf{
y}$-update, the minimization of \eqref{y_update_lagrange} with respect to $\mathbf{y}$ can be done by equating its first-order derivative to zero, that is
\begin{equation}
\begin{split}
    &-\boldsymbol{\gamma}
    + \mu\left(\mathbf{y}^{j+1} - \mathbf{z}^j + \mathbf{g}^j\right) \\
    &+ \rho\mathcal{H}\left(\mathcal{H}^H\left(\mathbf{y}^{j+1}\right) + \mathbf{S}^i - \boldsymbol{\Lambda} + \mathbf{W}^i\right)
     = \mathbf{0}.
\end{split}
\end{equation}

Reordering the terms and using the distributive property of linear maps lead to
\begin{equation}
\begin{split}
\rho\mathcal{H}\left(\mathcal{H}^H\left(\mathbf{y}^{j+1}\right)\right) &+ \mu\mathbf{y}^{j+1} \\
    &\hspace{-4.05em}= \boldsymbol{\gamma} - \rho\mathcal{H}\left(\mathbf{S}^i - \boldsymbol{\Lambda} + \mathbf{W}^i\right) + \mu\left(\mathbf{z}^j - \mathbf{g}^j\right).
\end{split}
\end{equation}

The operator $\mathcal{H}\left(\mathcal{H}^H\left(\mathbf{y}\right)\right)$ can be equivalently formulated as $\mathcal{H}\left(\mathcal{H}^H\left(\mathbf{y}\right)\right) = \mathbf{HH}^H\mathbf{y}$, where $\mathbf{H} \in \mathbb{C}^{K \times N^2}$ is defined as $\mathbf{H} = [\textrm{vec}(\mathbf{H}_1)^H; \hdots; \textrm{vec}(\mathbf{H}_K)^H]$ and $\textrm{vec}(\mathbf{H}_k)$ is the column-wise vectorization of $\mathbf{H}_k$. This reformulation allows for a closed-form update of $\mathbf{y}$ as
\begin{equation}
\begin{split}
    \mathbf{y}^{j+1} = \left(\rho\mathbf{HH}^H + \mu\mathbf{I}_K\right)^{-1}\Bigl(\boldsymbol{\gamma} &- \rho\mathcal{H}\left(\mathbf{S}^i - \boldsymbol{\Lambda} + \mathbf{W}^i\right) \\
    &+ \mu\left(\mathbf{z}^j - \mathbf{g}^j\right)\Bigr).
\end{split}
\label{inner_y_update}
\end{equation}
Note that the matrix inverse in \eqref{inner_y_update} needs to be computed only once before the start of the main (outer) ADMM algorithm. Moreover, the term $\boldsymbol{\gamma} - \rho\mathcal{H}\left(\mathbf{S}^i - \boldsymbol{\Lambda} + \mathbf{W}^i\right)$ is independent of the inner ADMM iterations so is computed once for every $\mathbf{y}$-update of the outer ADMM iterations. The remaining term is relatively cheap to compute, which results in a fast inner ADMM update.

As for the update of the auxiliary variable $\mathbf{z}$, problem \eqref{z_update} is equivalent to solving the following problem:
\begin{subequations}
\begin{align}
    \min_{\mathbf{z}} \quad &\norm{\mathbf{y}^{j+1} + \mathbf{g}^j - \mathbf{z}}_2^2 \\
    \textrm{s.t.} \quad &\mathbf{z} \geq \mathbf{0}.
\end{align}
\label{z_problem}%
\end{subequations}
It is clear from problem \eqref{z_problem} that the optimum $\mathbf{z}$-update is given by
\begin{equation}
    \mathbf{z}^{j+1} = \mathrm{max}\left(\mathbf{y}^{j+1} + \mathbf{g}^j, \mathbf{0}\right).
    \label{z_update2}
\end{equation}

\subsubsection{$\mathbf{S}$-Update}

We define the variable $\mathbf{X}^{i+1} = \boldsymbol{\Lambda} - \mathcal{H}^H\left(\mathbf{y}^{i+1}\right) - \mathbf{W}^i$. Problem \eqref{S_update} can be rewritten as
\begin{subequations}
    \begin{align}
        \min_{\mathbf{S}} \quad &\norm{\mathbf{S} - \mathbf{X}^{i+1}}_F^2 \\
    \textrm{s.t.} \quad &\mathbf{S} \succeq \mathbf{0}.
    \end{align}
\end{subequations}
The $\mathbf{S}$-update is then given by
\begin{equation}
    \mathbf{S}^{i+1} = \mathbf{X}_+^{i+1} \triangleq \mathbf{Q}_+^{i+1}\boldsymbol{\Sigma}_+^{i+1}(\mathbf{Q}_+^{i+1})^H,
\end{equation}
where $\boldsymbol{\Sigma}_+$ is a diagonal matrix with the non-negative eigenvalues of $\mathbf{X}^{i+1}$ and $\mathbf{Q}_+^{i+1}$ denotes a matrix with the corresponding eigenvectors as columns.

\subsubsection{Convergence Criteria}

The stopping criteria for the outer ADMM updates are chosen to verify the convergence of the primal and dual variables. To determine termination, the following two conditions must be satisfied at the $i^{th}$ iteration:
\begin{align}
    \frac{\bigl|\textrm{tr}\left(\mathbf{W}^i - \mathbf{W}^{i-1}\right)\bigr|}{\textrm{tr}\left(\mathbf{W}^i\right)} &< \epsilon_{\mathrm{dual}}, \\
    \frac{\norm{\mathbf{S}^i - \mathbf{S}^{i-1}}_F}{\norm{\mathbf{S}^i}_F} &< \epsilon_{\mathrm{prim}},
\end{align}
where $\epsilon_\mathrm{dual} > 0$ and $\epsilon_\mathrm{prim} > 0$ are predefined stopping conditions. We highlight that the stopping criterion for the dual variable $\mathbf{W}$ directly translates into the relative change in the original optimization objective. To avoid early termination due to stagnation, both conditions are deemed necessary to achieve the best possible outcome. As for the inner ADMM algorithm with cheap updates, we avoid computing stopping criteria to reduce the complexity per inner iteration. Instead, the algorithm is run for a fixed number of iterations $T$.

\subsection{Designing $\boldsymbol{\Lambda}$}

The SEA is a method to extract a near-optimal rank-1 solution through the elimination of the higher-rank solutions from the SDP. This method was introduced in our primal work \cite{your_globecom_paper}. The idea is to iteratively modify the optimization objective by introducing a penalty on the directions of the eigenvectors corresponding to the second largest eigenvalues of the higher-rank solutions outputted by the SDP in the previous iterations. The purpose of the design parameter $\boldsymbol{\Lambda}$ is to incorporate this penalty and improve the convergence speed of our SDP-ADMM implementation.

Since the trace and summation are linear operators, the objective of the original QoS problem \eqref{qos_problem} and the modified penalty of the SEA can be reformulated as
\begin{equation}
    \begin{split}
        \textrm{tr}\left(\mathbf{W}\right) + \sum_r\zeta_r\mathbf{u}_r^H\mathbf{W}\mathbf{u}_r &= \textrm{tr}\left(\mathbf{W}\right) + \textrm{tr}\Biggl(\sum_r\zeta_r\mathbf{U}_r\mathbf{W}\Biggr) \\
        &= \textrm{tr}\Biggl(\Biggl(\mathbf{I}_N + \sum_r\zeta_r\mathbf{U}_r\Biggr)\mathbf{W}\Biggr) \\
        &= \Biggl\langle\mathbf{I}_N + \sum_r\zeta_r\mathbf{U}_r, \mathbf{W}\Biggr\rangle,
    \end{split}
\end{equation}
where $\mathbf{U}_r = \mathbf{u}_r\mathbf{u}_r^H$. In this work, we make use of the linearity of the derivatives in the ADMM update equations by modifying the penalty proposed in \cite{your_globecom_paper}. The design parameter $\boldsymbol{\Lambda}$ is then set to
\begin{equation}
    \boldsymbol{\Lambda} = c\mathbf{I}_N + \sum_r\zeta_r\mathbf{U}_r,
\end{equation}
where $c \geq 1$ is a tunable weight set to improve the convergence speed towards the original optimization objective. Further, the penalty factors $\zeta_r$, $\forall r$ are modified and set equal to the corresponding eigenvalues of $\mathbf{U}_r$. We stress that this penalty is particularly beneficial for the ADMM algorithm updates. In this way, after a penalty is applied to a higher-rank output solution, the second strongest component of this higher-rank precoding matrix $\mathbf{W}$ is completely eliminated from the subsequent iterates $\{\mathbf{S}^i\}$ and $\{\mathbf{W}^i\}$ of our SDP-ADMM algorithm. We have observed that this modification improves the convergence speed and the numerical stability of the proposed SDP-ADMM algorithm compared to the previously proposed penalty term, however, it is not well-suited for higher-order methods.

\subsection{MMF Multicast Beamforming}

Since every UE in a multicast group must be able to decode the data, we assume a common SNR target for all the UEs, that is $\boldsymbol{\gamma} = \gamma_c$. However, we stress that the proposed algorithm is applicable to different SNR targets. The complete procedure is detailed in Algorithm \ref{alg1}.
Note that $P_T$ represents the power budget at the BS and $\kappa$ determines the elimination interval width, i.e., the bisection search interval for subsequent higher-rank solution elimination.

\begin{algorithm}
        \caption{MMF Multicast Beamforming via SDP-ADMM}
        \label{alg1}
        \noindent\textbf{Input:} The interval $[\gamma_\textrm{lo}, \gamma_\textrm{up}]$ that contains the optimum objective value of the MMF problem, $\gamma_\textrm{lo} = 0$ and $\gamma_\textrm{up} = \min\limits_{k} \frac{P_T\norm{\mathbf{h}_{k}}^2}{\sigma_{k}^2}$. Solution tolerance $\epsilon > 0$. ADMM penalty parameters $\rho$ and $\mu$, and stopping conditions $\epsilon_\mathrm{dual} > 0$ and $\epsilon_\mathrm{prim} > 0$. Number of inner ADMM iterations $T$. Initialize $\mathbf{y}^0 = \mathbf{z}^0 = \mathbf{g}^0 = \mathbf{0}$, $\mathbf{S}^0 = \mathbf{W}^0 = \frac{P_T}{N}\mathbf{I}_N$.
        
	\begin{algorithmic}[1]
            \REPEAT \label{bisection_start}
            \STATE Set $\gamma_c \leftarrow \left(\gamma_\textrm{lo} + \gamma_\textrm{up}\right) / 2$.
            \STATE Initialize $\mathbf{W}^0$ as a scaled version of the solution to the previous bisection iteration with the factor $\gamma_c/\gamma_\mathrm{prev}$.
            \STATE Solve the QoS problem \eqref{penalized_qos_problem} using the SDP-ADMM algorithm in Section \ref{SDP_ADMM} for the current $\gamma_c$.
            \IF{$\textrm{tr}\left(\mathbf{W}\right) > P_T$}
                \STATE Set $\gamma_\textrm{up} \leftarrow \gamma_c$.
            \ELSE
            \STATE Set $\gamma_\textrm{lo} \leftarrow \gamma_c$.
            \ENDIF
            \STATE Set $\gamma_\mathrm{prev} \leftarrow \gamma_c$.
            \UNTIL $\gamma_\textrm{up} - \gamma_\textrm{lo} < \epsilon$ \label{bisection_end}
            \STATE Set $r \leftarrow 1$.
            \STATE Set $\boldsymbol{\Lambda} \leftarrow c\mathbf{I}_N$.
            \WHILE{$\textrm{rank}\left(\mathbf{W}\right) \neq 1$}
            \STATE Set $\zeta_r \leftarrow$ Second strongest eigenvalue of $\mathbf{W}$.
            \STATE Set $\mathbf{U}_r \leftarrow \mathbf{u}_r\mathbf{u}_r^H$, where $\mathbf{u}_r$ is the corresponding eigenvector to $\zeta_r$.
            \STATE Update $\boldsymbol{\Lambda} \leftarrow \boldsymbol{\Lambda} + \zeta_r\mathbf{U}_r$.
            \STATE Set $\gamma_\textrm{lo} \leftarrow \kappa\gamma_c$. \label{kappa_interval}
            \STATE Repeat steps $\ref{bisection_start}$-$\ref{bisection_end}$.
            \STATE Set $r \leftarrow r + 1$.
            \ENDWHILE
\end{algorithmic}
\textbf{Output:} The near-optimal rank-$1$ solution.
\end{algorithm}

Two fundamental advantages of the proposed algorithm are worth noting. The first is that it does not require initialization with a feasible solution to the original QoS problem, which is necessary for algorithms that rely on the SCA technique (the state-of-the-art in multicast beamforming optimization).  Note that it might be hard to find an effective low-complexity solution to use as initialization, specially when the number of UEs grows large.
Another advantage is that the elimination procedure can perform the bisection search over the SNR target within a relatively small interval. This allows for a reduction in the computational time of the elimination procedure iterations as compared to the first solution (the conventional higher-rank SDP solution). The reason is that the upper bound of the bisection search interval will be the optimum $\gamma_c$ achieved in the previous bisection iteration, whereas the lower bound is chosen with a small separation away from it (Step \ref{kappa_interval}).

It is worth mentioning that at a given bisection iteration over the SNR target, the initial point for $\mathbf{W}$ is chosen as a scaled version of the solution to the previous bisection iteration. This provides a warm start to the ADMM algorithm and can greatly speed up the convergence to the optimal solution. In fact, if the QoS problem is solved exactly, i.e., a rank-1 solution is obtained directly from the SDP, then the scaled $\mathbf{W}$ is actually the optimal solution for the new SNR target. Otherwise, this scaled solution is fine-tuned with the ADMM algorithm for each step of the bisection search over the SNR target.

\section{Numerical Evaluation}

In this section, we use Monte Carlo simulations to verify the effectiveness of the proposed SDP-ADMM optimization procedure to solve the MMF multicast problem. We consider a massive MIMO system with a half-wavelength-spaced uniform linear array of $N = 36$ antennas deployed at the BS, serving an area of $750\,\textrm{m} \times 750\,\textrm{m}$. We assume the BS serves $K = \{15, 30\}$ UEs, that are randomly and uniformly distributed within the area of interest, using a single multicast transmission. The simulation parameters are summarized in Table \ref{params}. The channel between the BS and an arbitrary UE $k$ is modeled by correlated Rayleigh fading as $\mathbf{h}_{k} \sim \mathcal{N}_{\mathbb{C}}(\mathbf{0}, \mathbf{R}_{k})$, where $\mathbf{R}_{k} \in \mathbb{C}^{N \times N}$ represents the spatial correlation matrix, generated using the local scattering model in \cite{bjornson2017book}. The average channel gain, $\beta_{k} = \frac{1}{N} \hspace{1pt}\textrm{tr}\hspace{-1pt}\left(\mathbf{R}_{k}\right)$, is calculated using the 3GPP Urban Microcell model with correlated shadowing among the UEs. More precisely, the large-scale fading coefficients are given by
\begin{equation}
\beta_{k} = -30.5 - 36.7 \textrm{log}_{10}\left(\frac{d_{k}}{1\,\textrm{m}}\right) + F_{k} \hspace{2pt}\textrm{dB},
\label{pathloss}
\end{equation}
where $d_{k}$ represents the distance between the BS and UE $k$, and $F_{k} \sim \mathcal{N}\left(0, 4^2\right)$ denotes the shadow fading. The shadowing is correlated between the BS and different UEs as
\begin{equation}
\mathbb{E}\{F_{k}F_{i}\} = 
    4^22^{-\delta_{ki}/9\,\textrm{m}},
\end{equation}
where $\delta_{ki}$ is the distance between UE $k$ and UE $i$.

The SDP-ADMM penalty parameters are set to $\rho = 1$ and $\mu = 10^5$. The stopping conditions for the outer ADMM algorithm are set to $\epsilon_\mathrm{dual} = 10^{-5}$ and $\epsilon_\mathrm{prim} = 10^{-4}$, whereas the maximum number of iterations is set to $1000$. For the inner ADMM updates, the number of iterations is fixed to $T = 50$. The tunable weight defined for improving the convergence speed of the algorithm is set to $c = 5$. The stopping criterion for the bisection search is chosen to be $\epsilon = 0.1$ and the elimination interval uses $\kappa = 0.9$. We use the same platform for performing the simulations, a 4 core Intel(R) Core i5-10310U CPU with 1.7 GHz base frequency and 4.4 GHz turbo frequency. All programs are written in Matlab. The CDF curves are generated using $2000$ simulation samples.

\begin{table}
\vspace{0.1cm}
\begin{center}
\caption{Network simulation parameters.}
\begin{tabular}{ |c|c| }
\hline
Area of interest & $750\,\textrm{m} \times 750\,\textrm{m}$ \\
Bandwidth & $20$\,MHz \\
Number of BS antennas & $N = 36$ \\
Number of UEs & $K = \left\{15, 30\right\}$ \\
BS transmit power & $P_T = 40$\,W \\
DL noise power & $-94$\,dBm \\
\hline
\end{tabular}
\label{params}
\end{center}
\vspace{-1em}
\end{table}

\begin{figure}
\centering
\setlength{\abovecaptionskip}{0.33cm plus 0pt minus 0pt}
\includegraphics[scale=0.44]{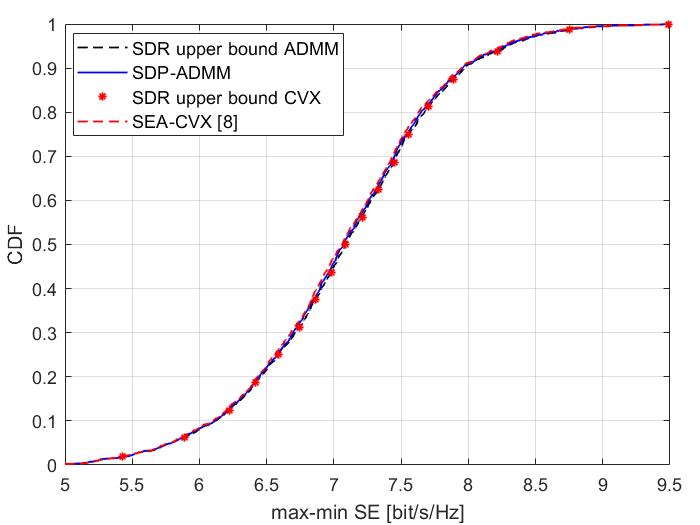}
\caption{CDF of the max-min SE for $K = 15$ UEs.}
\label{k15}
\end{figure}

\begin{figure}
\centering
\setlength{\abovecaptionskip}{0.33cm plus 0pt minus 0pt}
\includegraphics[scale=0.44]{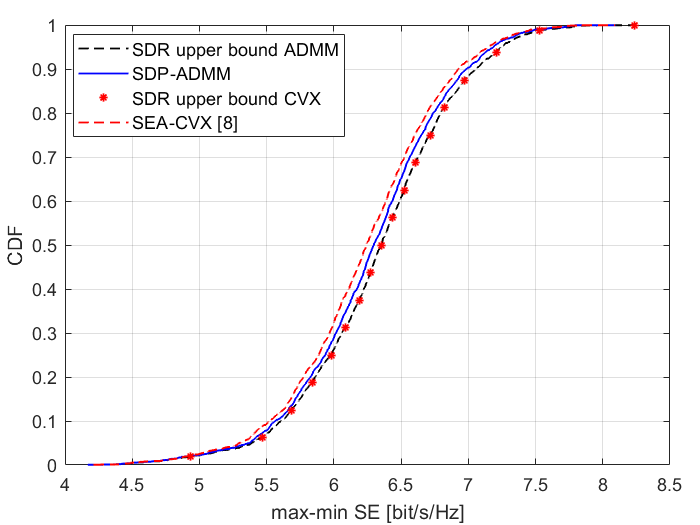}
\caption{CDF of the max-min SE for $K = 30$ UEs.}
\label{k30}
\end{figure}

Figs.~\ref{k15} and \ref{k30} plot the cumulative distribution function (CDF) of the max-min SE for the proposed SDP-ADMM algorithm for $K = 15$ and $K=30$ UEs, respectively. The state-of-the-art SEA algorithm \cite{your_globecom_paper}, which utilizes CVX \cite{cvx}, and the SDR upper bound are shown for comparison. It is clear that the proposed low-complexity SDP-ADMM algorithm exhibits similar SE performance to that of the SEA. Later, we will show the tremendous saving in computational time for the proposed SDP-ADMM algorithm. For $K = 15$ UEs, both the SDP-ADMM and SEA algorithms almost achieve the SDR upper bound. Remarkably, the proposed SDP-ADMM algorithm is able to achieve a small SE improvement over \cite{your_globecom_paper} for the denser scenario with $K = 30$ UEs, thanks to the refined penalty term that is specifically tailored for the SDP-ADMM implementation. We stress that this modified penalty term is not suitable for algorithms that rely on higher-order methods.

We highlight that we have selected our previously proposed SEA as a benchmark as it outperforms state-of-the-art SCA-based methods when standard solvers where used to solve the optimization problems, as demonstrated in \cite{your_globecom_paper}. The SE improvement over SCA-based methods stems from their use of a gradient descent approach, which ensures convergence only to a stationary point of the non-convex MMF problem. In contrast, the proposed algorithm begins with the optimal higher-rank solution obtained via SDR and iteratively reduces its rank by penalizing eigenvectors associated with the second-largest eigenvalues. This process continues until a near-optimal rank-1 solution is reached in the orthogonal subspace of these eigenvectors. At each iteration, the solution matrices are Hermitian positive semidefinite with distinct non-zero eigenvalues, ensuring that the penalty minimally impacts the optimality of subsequent iterations. As a result, this approach efficiently converges to a near-global optimal rank-1 solution to the NP-hard MMF multicast problem.

Table \ref{runtime} presents the average computational time of the proposed SDP-ADMM algorithm as compared to the benchmark. For the case of $K = 15$ UEs, the proposed algorithm is able to achieve more than $10$ times reduction in computational time as compared to the state-of-the-art SEA utilizing CVX and attains the same SE performance. As for the dense scenario of $K = 30$ UEs, the proposed algorithm brings an $88\,\%$ reduction in computational time while achieving a small improvement in SE.

\begin{table}
\vspace{0.1cm}
\begin{center}
\caption{Average computational time in seconds.}
\begin{tabular}{ |c|c|c|  }
\hline
$K$ & SDP-ADMM & SEA \cite{your_globecom_paper} \\
\hline
$15$ & $0.6$ & $8.6$ \\
$30$ & $2.1$ & $17.8$  \\
\hline
\end{tabular}
\label{runtime}
\end{center}
\vspace{-1em}
\end{table}

\section{Conclusions}

This paper addresses the MMF multicast optimization problem in a massive MIMO system. We develop a fast SDP-ADMM algorithm that can achieve a near-global optimal rank-1 beamforming solution to this NP-hard problem. The proposed solution relies on the SEA to extract a rank-1 solution from the output of the SDP. A novel problem reformulation and penalization scheme is proposed that is specifically tailored for our SDP-ADMM implementation. The distinction of the proposed algorithm is that it is applicable to a wide-class of optimization problems where SDR is utilized and a low-rank solution is desirable. Compared to standard SDP solvers, SDP-ADMM attains similar SE performance with about $10$ times reduction in computational time. Numerical results show the robustness of the proposed SDP-ADMM algorithm where it maintains its performance for different simulation settings without the need for ADMM parameter tuning.

\section*{References}
\renewcommand{\refname}{ \vspace{-\baselineskip}\vspace{-1.1mm} }
\bibliographystyle{ieeetr}
\bibliography{papercites}

\end{document}